\documentclass[english,11pt,aps,prd,nofootinbib,showkeys,preprint,floatfix]{revtex4-1}
\pdfoutput=1
\usepackage[utf8]{inputenc}
\usepackage[english]{babel}
\usepackage{amsmath}
\usepackage{amssymb}
\usepackage{graphicx}
\usepackage{bbold}
\usepackage{array,multirow} 
\usepackage[colorlinks=True]{hyperref}
\usepackage{lmodern}
\usepackage{subcaption}
\usepackage{color}
\usepackage{ulem}

\newcommand{\AddrUdeA}{%
 Instituto de F\'\i sica, Universidad de Antioquia,\\
 Calle 70 No. 52-21, Medell\'in, Colombia. }
 \newcommand{\AddrITM}{ Instituto Tecnol\'ogico Metropolitano, Facultad de Ciencias, Medell\'in, Colombia.} 

\begin{document}

\title{Singlet-doublet Dirac fermion dark matter from Peccei-Quinn symmetry}

\author{Robinson Longas}\email{robinson.longas@udea.edu.co}
\affiliation{\AddrUdeA} 
\author{Andr\'es Rivera} \email{afelipe.rivera@udea.edu.co}
\affiliation{\AddrUdeA}
\author{Cristian Ruiz} \email{cdavid.ruiz@udea.edu.co}
\email{cristianruiz6246@correo.itm.edu.co}
\affiliation{\AddrUdeA}
\affiliation{\AddrITM}
\author{David Suarez} \email{david.suarezr@udea.edu.co}
\affiliation{\AddrUdeA}


\begin{abstract}

Weakly Interacting Massive Particles (WIMPs) and axions are arguably the most compelling dark matter (DM) candidates in the literature. Here, we consider a model where the PQ symmetry solves the strong CP problem, generates radiatively Dirac neutrino masses, and gives origin to a multicomponent dark sector. Specifically, scotogenic Dirac neutrino masses arise at one-loop level. The lightest fermionic mediator acts as the second DM candidate due to a residual $Z_2$ symmetry resulting from the PQ symmetry breaking. The WIMP DM component resembles the well-known singlet-doublet fermion DM. While the lower WIMP dark mass region is usually excluded, our model reopens that portion of the parameter space (for DM masses below $\lesssim 100$ GeV). Therefore, we perform a phenomenological analysis that addresses the constraints from direct searches of DM, neutrino oscillation data, and charged lepton flavor violating (LFV) processes. The model can be tested in future facilities where neutrino telescopes search for DM annihilation into SM particles.
\end{abstract}

\maketitle

\section{Introduction}
\label{sec:intro}

Some evidence suggests the existence of Dark Matter (DM) and provides a way to study physics beyond the Standard Model (SM)~\cite{Zwicky:1937zza, Rubin:1970zza, Rubin:1980zd,clowebradacgonzalez2006, Refregier:2003ct, Tyson:1998vp}. However, the nature of DM remains obscure as its detection continues to be one of the big open problems nowadays. Many DM candidates have been proposed over the last years, particularly the Weakly Interacting Massive Particles (WIMPs)~\cite{Arcadi:2017kky}. A specific example is the singlet-doublet Dirac fermion presented in Ref~\cite{Yaguna:2015mva}. In that work, the author showed that the model's parameter space region was excluded for DM masses below $\sim 100$ GeV because of the large coupling between the DM particle and the $Z$ boson. Nonetheless, in this work, we show how this region is recovered.

On the other hand, the non-observation of CP violation in the Quantum Chromodynamics (QCD) Lagrangian represents one of the most active research topics in high-energy physics, both theoretically and experimentally. From a theoretical point of view, the absence of CP violation in the QCD Lagrangian is dynamically explained by invoking the Peccei Quinn (PQ) mechanism \cite{Peccei:1977hh}, which considers the spontaneous breaking of an anomalous global $\mathrm{U(1)}$ symmetry with the associated pseudo-Nambu-Goldstone boson, the (QCD) axion \cite{Weinberg:1977ma, Wilczek:1977pj}. The axion is a promising candidate for being the main component of DM of the Universe thanks to a variety of production mechanisms \cite{Sikivie:2006ni}; for instance, via the vacuum misalignment \cite{Preskill:1982cy, Abbott:1982af, Dine:1982ah}. Besides, it is remarkable that the physics behind the PQ mechanism can also explain other open questions such as neutrino masses \cite{Mohapatra:1982tc, Langacker:1986rj, Shin:1987xc, He:1988dm, Berezhiani:1989fp, Bertolini:1990vz, Ma:2001ac, delaVega:2020jcp}. For example, recent analysis that considers the PQ mechanism as responsible for the neutrino masses reveals that it is also possible to consistently provide a set of multicomponent scotogenic models with Dirac neutrinos \cite{Carvajal:2018ohk}. Specifically, in these scenarios, one-loop Dirac neutrino masses are generated through the $d = 5$ effective operator $\bar{L}\tilde{H}N_R\sigma$ \cite{Ma:2016mwh, Yao:2018ekp} once the axion field $\sigma$ develops a vacuum expectation value (VEV). In contrast, the contributions from the tree-level realizations of a dimension-five operator for neutrino masses are forbidden due to the charge assignment. As a further consequence of the PQ symmetry, the residual discrete symmetry stabilizes the lightest particle that mediates the neutrino masses. Since such a particle must be electrically neutral, this setup also accounts for a second DM species~\cite{Baer:2011hx, Dasgupta:2013cwa, Alves:2016bib, Chatterjee:2018mac, Ma:2017zyb}.

In this work, we enlarge the SM symmetry group with two new global symmetries:  a $\mathrm{ U(1)_{PQ} }$ and a $\mathrm{ U(1)_{L}}$ lepton number. Moreover, we add three right-handed singlets $\nu_{Ri}$ ($i=1,2,3$) corresponding to the SM neutrinos' right-handed partners. Additionally, we consider one $\mathrm{SU(2)_L}$ fermion singlet $N$, two $\mathrm{SU(2)_L}$ fermion doublets, $\eta$, $\psi$ and three scalar singlets $S_{\alpha}$ ($\alpha = 1,2,3$). Also, we consider a scalar singlet $\sigma$ that contains the axion field and one exotic chiral down-type quark $D$ that guarantees the realization of the hadronic KSVZ axion model \cite{Kim:1979if, Shifman:1979if}. This model was cataloged as T1-2-B in Ref~\cite{Carvajal:2018ohk}, where the spontaneous symmetry breaking of the PQ symmetry provides a mechanism for one-loop Dirac neutrino masses. We perform a phenomenological analysis of the model by determining the viable parameter space from direct detection (DD) experiments, lepton flavor violating (LFV) processes, DM relic density, neutrino physics, and indirect detection searches in neutrino telescopes. The model easily satisfies these constraints. Also, a considerable portion of its parameter space will be tested by future experiments. The WIMP DM component is a mixture between the singlet and the doublet Dirac fermions that resembles the well-known singlet-doublet fermion DM~\cite{Yaguna:2015mva, Restrepo:2019soi}. However, we show that since new DM annihilation channels lead to the correct relic density, the lower mass region (for DM mass below $\lesssim 100$ GeV) is reopened. 

This paper is organized as follows. Section \ref{sec:model} describes the model and its constraints. 
Section \ref{sec:DM_pheno} shows the DM phenomenology. Section \ref{sec:results} contains numerical analysis and discusses the results. Finally, section \ref{sec:conclusions} concludes.

\section{The model}
\label{sec:model}

We add to SM the right-handed partners of the neutrinos, the singlets  $\nu_{Ri}$ ($i=1,2,3$). Additionally, we consider as new particle content of the model: one $\mathrm{SU(2)_L}$ fermion singlet $N$, two $\mathrm{SU(2)_L}$ fermion doublets, $\eta$, $\psi$ and three scalar singlets $S_{\alpha}$ ($\alpha = 1,2,3$). Those fields are required for the one-loop realization of neutrino masses. Also, we consider a scalar singlet $\sigma$ that contains the axion field and one exotic chiral down-type quark $D$ that guarantees the realization of the hadronic KSVZ axion model \cite{Kim:1979if, Shifman:1979if}. The  particle content of the model and the charge assignments under the global symmetries, $\mathrm{U(1)_L}$ and $\mathrm{U(1)_{PQ}}$, are displayed in Table \ref{tab:particle-content-Dirac}. Notice that in this model, the SM model leptons have PQ charges, and the SM Higgs and the ordinary quarks are neutral under the global symmetries. 

\begin{table}
 \centering
 \begin{tabular}{|c|c|c|c|c|c|c|c|c|c|c|c|}
 \hline
  & $L_i$ & $e_{Ri}$ & $\nu_{Ri}$ & $N$& $\psi$ & $\eta$ & $S_{\alpha}$ & $D_L$ & $D_R$ & $\sigma$  \\
 \hline \hline
 ${\rm U(1)_L}$ & 1 & -1 & 1 & 1 & 1 & 1 & 0 & 0 & 0 & 0 \\
 \hline
 ${\rm U(1)_{PQ}}$  & 2 & -2 & 0 & 1 & 1 & 3 & 1 & 1 & -1 &2  \\
 \hline \hline
 ${\rm Z_2}$  & + & + & +& - & -& - & - & - & - & + \\
 \hline
 \end{tabular}
 \caption{ Particle content of the model with its lepton and PQ charge assignments. We also show the transformation under the remnant $Z_2$ symmetry.} 
 \label{tab:particle-content-Dirac}
 \end{table}

The most general Lagrangian invariant under such symmetries contains the following terms,
\begin{align}
\label{eq:ModelLagrangian}
\mathcal{L}\supset &~ M_N \overline{N}N +M_{\psi} \overline{\psi} \psi + M_{\eta} \overline{\eta}\eta + \left [~ \lambda_1  \overline{\psi_L} \eta_R \sigma^{\ast}  + \lambda_2 \overline{\eta_L} \psi_R \sigma + \kappa_1 \overline{\psi_L} \tilde{H} N_R + \kappa_2\overline{\psi_R} \tilde{H} N_L \right.  \nonumber \\
& \left. +~ h_{i\alpha} \overline{\eta_R} L_i S_{\alpha}  + f_{i\alpha} \overline{\nu_R}_i N_L S_{\alpha}^{\ast} + y_Q \overline{D_L}D_R\sigma  + \rm{h.c} ~\right]  - \mathcal{V}\left(H,S_{\alpha},\sigma \right)  \;,
\end{align}
where $\Tilde{H} = i\sigma_2 H^{\ast}$ and $\mathcal{V}\left(H,S_{\alpha},\sigma \right)$ is the scalar potential. $h_{i\alpha}$ and $f_{i\alpha}$ are the Yukawa couplings relevant for neutrino masses. The Yukawa couplings $\lambda_i$ ($i=1,2$) control the interactions between the axion field and the new fermions.  When PQ symmetry is broken, they provide mass terms for dark sector fermions. On the other hand, the Yukawa couplings $\kappa_i$ ($i=1,2$) mix the fermion singlet and the fermion doublet states. For this reason, they are crucial for the WIMP DM phenomenology~\cite{Yaguna:2015mva, Restrepo:2019soi}.

The scalar potential $\mathcal{V}\left(H,S_{\alpha},\sigma \right)$ reads as:
\begin{align}
\label{eq:scalarpotential}
\mathcal{V}\left(H,S_{\alpha},\sigma \right)= &~-\mu_1^2 |H|^2+\lambda_H|H|^4 +\mu_{S_{\alpha}}^2 |S_{\alpha}|^2+\lambda_{S_\alpha}|S_{\alpha}|^4
- \mu_{\sigma}^2 |\sigma|^2+\lambda_{\sigma}|\sigma|^4  \nonumber \\ & +\lambda_{HS_{\alpha}}|H|^2|S_{\alpha}|^2\,.
\end{align}
Here, we neglect the terms $|H|^2|\sigma|^2$ and $|S_{\alpha}|^2|\sigma|^2$  by rendering the respective quartic couplings small enough to avoid the scalar mixing between $H$ and $\sigma$. This mixing plays only a role for inflation signatures~\cite{Ballesteros:2016xej, Ringwald:2020vei}. 

We demand the stability of the scalar potential. It is bounded from below by imposing the copositivity conditions~\cite{Kannike:2012pe,10.2307/2324036}:
\begin{align}
\label{eq:vacuu-stability}
 & \lambda_H\geq0,\ \ \ \lambda_\sigma\geq0,\ \ \ \lambda_{S_\alpha}\geq0, \nonumber \\
  &-\lambda_{HS_{\alpha}}+2\sqrt{\lambda_H\lambda_{S_\alpha}}\geq0, \ \ \ \sqrt{\lambda_H\lambda_\sigma}\geq0,\ \ \ \sqrt{\lambda_{S_\alpha}\lambda_\sigma}\geq0, \\
  & \lambda_{HS_\alpha}+\sqrt{\lambda_\sigma}+2\Bigg[\sqrt{\lambda_H\lambda_{S_\alpha}\lambda_\sigma}+\sqrt{\Big(\lambda_{HS_\alpha}+2\sqrt{\lambda_H\lambda_{S_\alpha}}\Big)\lambda_\sigma\sqrt{\lambda_H\lambda_{S_\alpha}}}\Bigg]\geq0\,,\;( \alpha=1,2,3)\,, \nonumber
\end{align}
together with $\mu_1^2\geq0$, $\mu_{\sigma}^2\geq0$ and $\mu_{S_\alpha}^2\geq0$ ($\alpha=1,2,3$). 

We write the scalar fields as:
\begin{align}\label{eq:scalarfields}
    \sigma=\frac{1}{\sqrt{2}}\left(\rho + v_{\sigma}\right)e^{ia/v_{\sigma}},\hspace{0.5cm}
    H=\begin{pmatrix}
    0\\
    \frac{v+h}{\sqrt{2}}
    \end{pmatrix}, \hspace{0.5cm} S_{\alpha}\,,
\end{align}
where $\rho$ stands for the radial component of the field $\sigma$ whose mass is set by the PQ symmetry breaking scale $v_{\sigma}$, whereas $a$ is the CP-odd component of $\sigma$ scalar that corresponds to the QCD axion field. In our notation, $h$ is the SM Higgs boson with a VEV $v\sim 246 $ GeV. Also, at low energies, the scalar spectrum comprises three $Z_2$-odd scalars that are assumed to be in the diagonal basis $(S_1, S_2, S_3)$, 
\begin{align}
\label{eq:z2oddscalarssimpler}
    m_S^2 = 
    \begin{pmatrix}
    \mu_{S_1}^2 + \frac{\lambda_{HS_1}}{2}v^2 & 0  & 0 \\
    0 & \mu_{S_2}^2 + \frac{\lambda_{HS_2}}{2}v^2  & 0 \\  
    0 & 0 & \mu_{S_3}^2 + \frac{\lambda_{HS_3}}{2}v^2   
    \end{pmatrix}\,.
\end{align}
Although the lightest scalar state could be a proper DM candidate, we focus on fermionic DM.

On the other hand, since both scalars $\sigma$ and $H$ acquire VEV, the Yukawa couplings: $\lambda_j, \kappa_j$ ($j=1,2$) in Eq. \eqref{eq:ModelLagrangian}  mix the fermion singlet and doublets. $h_{i\alpha}$ and $f_{i\alpha}$ represent pure interaction terms that generate Dirac neutrino mass terms (also, they may affect LFV processes).  After symmetry breaking, the mass spectrum contains three neutral and two charged Dirac fermions. In the basis $\Sigma_L = \left(\eta^-_L \;\;\psi^-_L\right)^T $, $\Sigma_R = \left( \eta^-_R \;\; \psi^-_R \right)^T $ we have the mass matrix: 

\begin{align}\label{eq:YukASBCharged}
{\bf M}_{\Sigma^\pm} = 
\begin{pmatrix}
M_{\eta} & \frac{\lambda_2 v_{\sigma}}{\sqrt{2}}  \\
\frac{\lambda_1 v_{\sigma}}{\sqrt{2}} & M_{\psi}
\end{pmatrix} \;.
\end{align}

It follows that the charged fermion spectrum of this model comprises two states $\chi_{1,2}^{\pm}$ with masses $m_{\chi_{1,2}^{\pm}} = \frac{1}{2}\left[ M_{\psi} + M_{\eta} \mp \sqrt{\left( M_{\psi} - M_{\eta}\right)^2 + 2\lambda_1\lambda_2 v_{\sigma}^2}\right]$, where $m_{\chi_2}^{\pm}>m_{\chi_1}^{\pm}$. 

For the neutral sector, in the basis $\Xi_L = \left(N_L\;\; \eta^0_L \;\;\psi^0_L\right)^T  $, $\Xi_R = \left(N_R\;\; \eta^0_R \;\;\psi^0_R\right)^T  $, we have the mass matrix: 
\begin{align}\label{eq:YukASB}
{\bf M}_{\Xi^0} =
\begin{pmatrix}
M_N & 0 & \frac{\kappa_2 v}{\sqrt{2}} \\
0 & M_{\eta} & \frac{\lambda_2 v_{\sigma}}{\sqrt{2}} \\
\frac{\kappa_1 v}{\sqrt{2}} & \frac{\lambda_1 v_{\sigma}}{\sqrt{2}} & M_{\psi}
\end{pmatrix}\;,
\end{align}
that is diagonalized by a biunitary transformation, $\chi_L = V_L \Xi_L$ and $\chi_R = V_R \Xi_R$. Then, we obtain three neutral Dirac fermion mass eigenstates where their masses are given by:
\begin{align} \label{eq:M0rotation}
m_{\chi_i}^{\mathrm{diag}} \equiv \mathrm{diag} \left(m_{\chi_1},m_{\chi_2},m_{\chi_3} \right)  = V_L{\bf M}_{\Xi^0}V_R^{\dagger}\,.
\end{align}

Our WIMP DM candidate is the lightest neutral state\footnote{ From now on, we set $\chi\equiv \chi_1$ to represent the fermionic DM candidate.}, $\chi_1$. Notice that the spectrum is quite similar to the one shown in Ref. \cite{Restrepo:2019soi}; but, in our model, there are two $\mathrm{SU(2)_L}$ vector-like fermions and a mass term that comes from the breaking of the PQ symmetry. For this reason, we have three neutral fermionic dark particles instead of two.

\subsection{Neutrino masses}
\label{sec:neutrinomasses}

\begin{figure}[th]
  \includegraphics[scale=0.8]{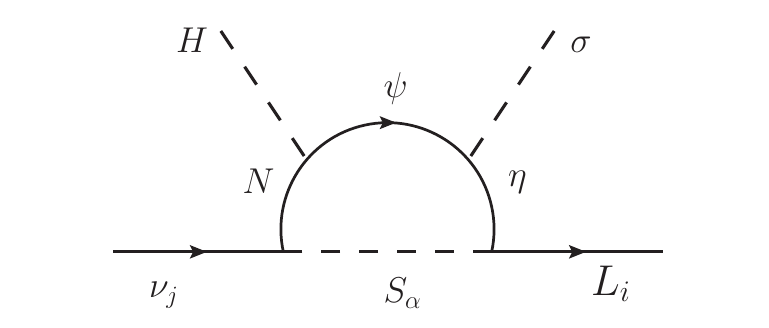}
  \caption{ Feynman diagram for one-loop Dirac neutrino masses in the interaction basis.} 
  \label{fig:Nmasses}
\end{figure}

The Yukawa Lagrangian in Eq. \eqref{eq:ModelLagrangian}  leads to one-loop neutrino masses via the couplings $f_{i\alpha}$ y $h_{i\alpha}$ as displayed in Fig. \ref{fig:Nmasses}. In the low-momentum limit, the neutrino mass matrix reads as:
\begin{align}\label{eq:neutrinomasses}
M^{\nu}_{ij} = \sum_{\alpha=1}^3 h_{i\alpha}\Lambda_{\alpha}f_{j\alpha} \Longleftrightarrow M^{\nu} = h\Lambda f^T\,,
\end{align}
where the loop integral factor $\Lambda_{\alpha}$ is:
\begin{align}\label{eq:Lambdafactor}
\Lambda_{\alpha} = \frac{1}{16\pi^2}\sum_{l=1}^3 \left(V_{R}\right)^\ast_{l2}~ \left(V_{L}\right)^\ast_{l1}~ m_{\chi_l}\times \left[ \frac{ m_{\chi_l}^2\ln\left(m_{\chi_l}^2\right) - m_{S_{\alpha}}^2\ln\left( m_{S_{\alpha}}^2\right)}{m_{\chi_l}^2-m_{S_{\alpha}}^2}\right]\,,
\end{align}
and the convergence of such a loop factor is guaranteed by the identity:
\begin{align}\label{eq:identityloop}
\sum_{l=1}^3 \left(V_{R}\right)_{l2}~ \left(V_{L}\right)_{l1}~ m_{\chi_l} = 0 \,.
\end{align}

The Dirac neutrino mass matrix in Eq. \eqref{eq:neutrinomasses} is diagonalized via a biunitary transformation $m= U^{\dagger} M^{\nu} V$, where $U$ and $V$ are unitary matrices and $m = {\rm diag}(m_1,m_2,m_3)$ is the diagonal matrix that contains three (or two) non-zero eigenvalues. They correspond to the masses of the neutrino mass eigenstates. In the basis where the charged lepton mass matrix is diagonal,  the unitary matrix $U$ is identified with the Pontecorvo-Maki-Nakagawa-Sakata (PMNS) matrix \cite{Maki:1962mu}, whereas $V$ is assumed diagonal without lost of generality. Furthermore, we further simplify our analysis by imposing one massless neutrino, $m_1 =0$ in the case of normal hierarchy (NH) and $m_3=0$ in the case of inverted hierarchy (IH). The Yukawa couplings $h_{i\alpha}$ are written in terms of $f_{i\alpha}$ and the neutrino observables as (see appendix \ref{sec:Yukawas} for details): 
\begin{align}
\label{eq:yukawaf}
h=U_{{\rm PMNS}}\sqrt{D}R\sqrt{D}\left(f^T\right)^{-1}\Lambda^{-1}\;,
\end{align}
where,
\begin{align}
    R = \left\{ \begin{array}{lcc}
\begin{pmatrix}
0&0&0\\
0&1&0\\
0&0&1
\end{pmatrix} &   {\mathrm{for}}  & {\mathrm{NH}}  \;,
\\ \\ \begin{pmatrix}
1&0&0\\
0&1&0\\
0&0&0
\end{pmatrix} &   {\mathrm{for}}   & {\mathrm{IH}} \;,
\end{array}
\right.
\end{align}
and: 
\begin{align}
\sqrt{D} = \left\{ \begin{array}{lcc}
{\rm diag}(\sqrt{v},\sqrt{m_2},\sqrt{m_3}) &   {\mathrm{for}}  & {\mathrm{NH}}  \,,
\\ {\rm diag}(\sqrt{m_1},\sqrt{m_2},\sqrt{v}) &   {\mathrm{for}}   & {\mathrm{IH}} \,.
\end{array}
\right.
\end{align}
where $v$ is some non-vanishing arbitrary energy scale.

\subsection{Standard Model Constraints} 

Current and coming experiments impose constraints on SM observables with sensitivity to new physics. One of them is the decay of SM Higgs boson into invisible particles. In the present model, the Higgs of the SM, $h$, interacts with the singlet scalars $S_{\alpha}$ through the scalar couplings $\lambda_{HS_{\alpha}}$ and with the neutral Dirac fermion via the Yukawa couplings $\kappa_1$ and $\kappa_2$. Therefore, in the low mass regime for DM masses lighter than $m_h/2$, the SM Higgs could decay to a fermion DM pair\footnote{We will not take into account the invisible Higgs decay into a scalar pair because we consider $\lambda_{HS_{\alpha}} \sim 10^{-4}$ and consequently the corresponding amplitude width is negligible.}. The partial decay width is given by:
\begin{align}\label{eq:InvisibledecayW}
\Gamma \left(h \rightarrow \chi \chi\right) \simeq  & \frac{3}{32\pi m_h} \sqrt{1 - \frac{m_{\chi}^2}{m_h^2}} \left[ |V_{R_{11}}|^2|V_{L_{13}}|^2  \left( m_h^2 - 2 m_{\chi}^2\right) \left (\kappa_1^2 + \kappa_2^2 \right)\right. \nonumber \\
& \left. - 2 m_{\chi}^2 \kappa_1^2\kappa_2^2\left( V_{R_{13}}^{\ast}V_{R_{11}}V_{L_{11}}^{\ast}V_{L_{13}} +  V_{L_{13}}^{\ast}V_{L_{11}}V_{R_{11}}^{\ast}V_{R_{13}} \right)\right]\,,
\end{align}
where $m_h$ and $m_{\chi}$ are the SM Higgs and the DM masses respectively. The branching ratio for the Higgs invisible decay is given by: 
\begin{align}\label{eq:branchingHinvisible}
\mathcal{B}_{h\rightarrow\mathrm{inv}} = \frac{\Gamma \left(h \rightarrow \chi \chi\right)}{\Gamma_{h,\mathrm{SM}} + \Gamma \left(h \rightarrow \chi \chi\right)}\,,
\end{align}
where $\Gamma_{h,\mathrm{SM}} \sim$ 4.1 MeV is the total decay width of the Higgs boson in the SM. The current limit from the Higgs invisible decay width is given by ATLAS $\mathcal{B}_{h\rightarrow\mathrm{inv}}<$ 0.13 \cite{ATLAS:2020cjb} and CMS $\mathcal{B}_{h\rightarrow\mathrm{inv}}<$ 0.19 \cite{CMS:2018yfx}. The prospects from the High Luminosity LHC (HL-LHC), $\mathcal{B}_{h\rightarrow\mathrm{inv}}<$ 0.019, and the Future Circular Colliders (FCC), $\mathcal{B}_{h\rightarrow\mathrm{inv}}<$ 0.00024, are summarized in Ref \cite{deBlas:2019rxi}. These limits will be imposed in section \ref{sec:DM_pheno} to constrain the parameter space of this model.

On the other hand, current experiments constrain the $Z$ Gauge boson decay into invisible states. In our model, the decay rate for $Z$ boson into DM fermions, when the DM fermions are lighter than $m_Z/2$, is given by~\cite{Dreiner:2008tw}:
\begin{equation}
\Gamma\left(Z\rightarrow\chi\chi\right)=\sum^3_{l=1}\left|(V_{R})_{l1}\right|^2\left|(V_{L})_{l1}\right|^2\frac{gm_Z}{96\pi\operatorname{c}^2_W}\left(1-\frac{m^2_{\chi_l}}{m^2_Z}\right)^{\frac{3}{2}}\,.    
\end{equation}
Current experiments show an upper bound for decay of $Z$ boson into invisible states with a decay width~\cite{ParticleDataGroup:2022pth}:
\begin{equation}
\Gamma\left(Z\rightarrow\text{invisible}\right)=499.0\pm{1.5}\, \text{MeV}\,.
\end{equation}
However, Section \ref{sec:results} shows that the DM particle is mostly a singlet-like fermion in the low mass regime, and then this observable remains within the experimental limit.

\begin{figure}
\includegraphics[scale=0.9]{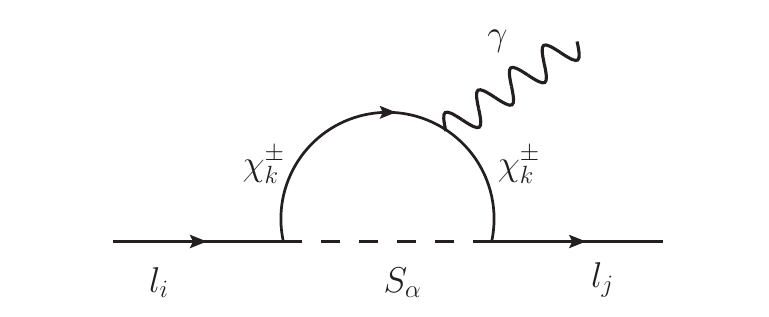}
  \caption{Feynman diagram that contributes to the $\mu \to e \gamma$ LFV process, where $i$ and $j$ are flavor indices.}
  \label{fig:LFV}
\end{figure}

On the other hand, LFV processes are susceptible to the contributions of new physics. Although the Diracness of neutrino masses is compatible with the conservation of the total lepton number, family lepton number violation is unavoidable due to neutrino oscillations. In this model, LFV processes that involve charged leptons are controlled by the Yukawa coupling $h_{i\alpha}$ ( see Lagrangian in Eq. ~\eqref{eq:ModelLagrangian}). One of the most restrictive LFV processes is the radiative muon decay $\mu \to e \gamma$, shown in Fig.~\ref{fig:LFV} for $i = 2$ and $j = 1$. 

Following  Ref.~\cite{Lavoura:2003xp}, we compute this branching ratio,
\begin{align}
 \label{eq:muegamma}
 \mathcal{B}(\mu\to e \gamma)
=\dfrac{3\,\alpha}{32\pi G_F^2}\sum_{\alpha=1}^3
\sum_{k=1}^2\left|h_{2\alpha}F(x)h^*_{1\alpha }(U_{\Sigma_R})_{1k}^2
\right|^2\,,
\end{align}
where $G_F$ is the Fermi constant and $\alpha = e/4\pi$ is the fine structure constant and $U_{\Sigma_R}$ is the mixing matrix for right-handed charged fermions . The loop function $F$ is given by, 
\begin{align}
\label{eq:Fx}
 F(x)=\left(\frac{x^3-6x^2+3x+2+6x\ln(x)}{6(x-1)^4}\right)\,,
 \end{align}
where $x=\left(\frac{m_{\chi_k}^{\pm}}{m_{S_\alpha}}\right)^2$.

One final observable susceptible to new physics, is the Peskin-Takeuchi, $S, T$, and $U$ parameters that render radiative corrections to masses of electroweak bosons. Because we consider a minor mixing angle between charged fermions, the $S$, $T$, and $U$ parameters in our model are controlled by the charged fermions masses that satisfy the experimental bounds for all observables.  

\section{Dark matter phenomenology}
\label{sec:DM_pheno}

In this model, there are two DM candidates: the axion $(a)$ that is a natural candidate after the PQ symmetry breaking mechanism and the WIMP candidate that is the lightest $Z_2$-odd state, {\it i.e}, either the scalar $S_1$ or the fermion $\chi$. Non-perturbative QCD effects determine the relic density of axions. At temperatures above the QCD critical temperature, $\Lambda_{\mathrm QCD}\sim 160$  MeV, the chiral symmetry is restored, then the axion is massless. The misalignment angle $\theta_a \equiv a/v_{\sigma}$ parameterizes the corresponding axion field. Later, as the temperature of the primordial plasma falls below the hadronic scale, the axion becomes a pseudo-Nambu-Goldstone boson and develops a mass due to non-perturbative effects \cite{Peccei:1977hh, Peccei:1977ur, GrillidiCortona:2015jxo}. The axion field oscillates around its mean value when its mass exceeds the Hubble expansion rate. These coherent and spatially uniform oscillations correspond to a coherent state of nonrelativistic axions where they behave as a cold DM fluid since their energy density scales as ordinary matter \cite{Arias_2012, DiLuzio:2020wdo, Nelson:2011sf}. 

The value of the component of the relic density provided by the axion strongly depends on the cosmological scenario. In other words, it is different if the PQ symmetry is broken after or during inflation. In a post-inflationary phase, the expected energy density depends on the misalignment angle and the scale of the PQ symmetry breaking $v_\sigma$, so that $\theta_a$ takes different values in different patches of the Universe, an average is $\theta_a\sim\pi^2/3$. In this case, possible topological defects such as axion strings and domain walls contribute to the axion energy density \cite{Marsh:2015xka, Sikivie:2006ni, Sikivie:2009qn, Arias_2012, DiLuzio:2020wdo}. Nevertheless, when the PQ symmetry is broken before the end of inflation, the topological defects are absent and the misalignment mechanism renders the axion relic density. In this scenario, the axion DM abundance is given by~\cite{Abbott:1982af, Ellis:2018dmb}:
\begin{align}
\label{eq:axionDMabunce}
 \Omega_ah^2\approx 0.18\theta_a^2\left(\frac{v_\sigma}{10^{12} \textrm{ GeV}}\right)^{1.19}\,.
\end{align}

From Eq. \eqref{eq:axionDMabunce} follows that the axion can compose the total amount of the DM constituent if $v_{\sigma}\sim 10^{12}$ GeV for $\theta_a\sim \mathcal{O}(1)$. Under this premise, the axion window becomes $m_a \sim (1-10)$~$\mu$eV. Nevertheless, the axion could give a subdominant contribution to the relic DM abundance for lower values. Thus, it allows a multicomponent DM scenario. In this work, the WIMP component dominates the relic abundance of DM in the Universe and the axion field plays a role in the generation of a Dirac mass term for neutrinos. The phenomenological study where the axion and the WIMP components are relevant was performed in Ref.~\cite{Carvajal:2021fxu}.

In addition to the axion, this model leads to a second DM candidate because the lightest $Z_2$-odd state is stable. This can be accomplished by either the scalar $S_1$ or the Dirac fermion $\chi$. If the scalar $S_1$ is the lightest state, the DM phenomenology resembles the Majorana version presented in Ref.~\cite{Restrepo:2015ura}. Conversely, for Dirac fermion DM, the DM phenomenology is given by the mixing between the singlet and the doublet fermion states studied in Refs. \cite{Yaguna:2015mva, Restrepo:2019soi}. If the lightest Dirac fermion in the dark sector is mainly singlet, the DM does not annihilate efficiently in the early Universe then its present abundance is greater than currently observed. On the other hand, if the DM candidate is mainly doublet, the correct relic abundance is recovered only for $m_{\chi} \sim 1$ TeV. In the singlet-doublet scenario, it was shown in Ref.~\cite{Yaguna:2015mva} that the region $100\, \text{GeV} \lesssim m_{\chi}\lesssim 750$ GeV is still available thanks to coanihiliations between fermions in the dark sector. However, the sizeable couplings between the DM candidate and the $Z$ boson, exclude the low mass region, $m_{\chi}\lesssim 100$ GeV. One of the features of the model presented here is that the region of the parameter space below $m_{\chi}\lesssim 100 $ GeV is recovered due to new DM annihilation channels. 

\subsection*{WIMP light dark matter window}
\label{sec:ligh_masswindow}

Unitarity condition imposes an upper bound of $\sim 340$ TeV on the DM mass of thermal relics \cite{Griest:1989wd}.  However, for low masses, limits are not so easily applied. Refs. \cite{Lee:1977ua, Hut:1977zn} show that fermion DM masses below a few GeV's are usually ruled out because the DM overcloses the Universe. Nevertheless, this limit can be evaded by considering new light mediators in thermal equilibrium with the DM candidate~\cite{Boehm:2003hm}\footnote{In this model, the $S_{\alpha}$ $(\alpha =1,2,3)$ scalars are the light mediators, which in the low mass window have masses around $\sim 1$ GeV.}. 

On the other hand, distortion to the CMB spectrum caused by energy injection to the primordial plasma, when DM couples to electrons, imposes a lower bound for the DM candidate mass of $\sim 10$ GeV. Nonetheless, this limit could be evaded if DM annihilates to neutrinos as well as electrons\footnote{This is achieved by the Yukawa term $f_{i\alpha} \nu_i^C N S_{\alpha}$  in the Lagrangian \eqref{eq:ModelLagrangian}.}. Conversely, if the DM is in thermal equilibrium with neutrinos, electrons, and photons, and if it decouples when it is non-relativistic, there is a change in the effective number of neutrinos $N_{\mathrm {eff}}$. A lower limit for the mass of the Dirac fermion DM of $m_{\chi}\gtrsim 10$ MeV, was given in Ref. \cite{Boehm:2013jpa}.

\begin{figure}[t!]
 \includegraphics[scale=0.9]{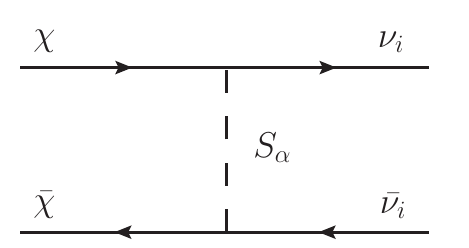}
   \caption{Feynman diagram for DM annihilation into a neutrino pair. }
   \label{fig:DMtoneutrinos}
 \end{figure}

It is well-known that for DM masses below $m_{\chi} \lesssim$ 5 GeV, the direct detection searches are not sensitive to scattering between DM candidates and nuclei, even for the expected future experiments as DARWIN~\cite{DARWIN:2016hyl}. For this reason, we consider indirect searches that look for DM annihilation into a neutrino pair as shown in Fig. \ref{fig:DMtoneutrinos}, and therefore extra neutrino flux is produced and detected by neutrino telescopes. There, neutrinos interact with the nuclei in the detector. After that, an electromagnetic signal is produced. The signal events are compared with measurements at the Super-Kamiokande (SK), Hyper-Kamiokande (HK), Deep  Underground Neutrino Experiment (DUNE), and Jiangmen Underground Neutrino Observatory (JUNO). Such experiments derive an upper limit on the annihilation cross section of the DM into neutrinos (see Ref~\cite{Arguelles:2019ouk} for a review).

The expected contribution for the neutrino flux from DM annihilation in the Milky Way halo is given by \cite{Arguelles:2019ouk, Okawa:2020jea}:
\begin{align}\label{eq:neutrinoflux}
\frac{d\Phi_{\nu \bar{\nu}}}{dE_{\nu}} = \frac{1}{16\pi m_{\chi}^2} \sum_i \langle \sigma v \rangle_i k \frac{dN_i}{dE_{\nu}}J\left(\Omega\right)\,,
\end{align}
where $k$ gives the electron-neutrino flavor factor, $\langle \sigma v \rangle_i$ stands for the annihilation cross section into a final state $i$, $dN_i/dE_{\nu}$ is the neutrino spectral function for the final state $i$ and $J\left(\Omega\right)$ represents the astrophysical $J$-factor. In the galactic coordinates, $(b,l)$, the $J$-factor can be expressed as,
\begin{align}\label{eq:Jfactor}
J = \int d\Omega \int_{\mathrm{l.o.s}} \rho^2\left(r\right)dr\, ,
\end{align}
where $\rho\left(r\right)$ is the DM density profile in the galactic halo. We consider here the Navarro-Frenk-White (NFW) profile.

In this model, the thermally averaged DM annihilation cross section times the velocity reads as:
\begin{align}
    \langle \sigma v \rangle \simeq  \sum_{\alpha=1}^3\sum_{i=1}^3\dfrac{1}{32\pi}\dfrac{( m_{\nu_i}^2+m_{\chi}^2)}{( m_{S_\alpha}^2+m_{\chi}^2-m_{\nu_i}^2)^2}
    \sqrt{1-\dfrac{m_{\nu_i}^2}{m_{\chi}^2}}\left(f_{i\alpha}^2+h_{i\alpha}^2\right)^2\,,
\end{align}
where $m_{\nu_i}$ $i=1, 2, 3$ are the masses of the neutrinos, $m_{\chi}$ is the DM mass, $m_{S_\alpha}$ are the masses for the scalar mediators, $f_{i\alpha}$ and $h_{i\alpha}$ are the Yukawa couplings defined in Eq. \eqref{eq:ModelLagrangian}. 
\begin{table}[t]
\begin{center}
 \begin{tabular}{||c||}
  \hline \hline \rule[0cm]{0cm}{.25cm}
$ 10^{-2}\,{\rm GeV}\leq M_N, M_{\psi}, M_{\eta} \leq 2\,{\rm TeV} $ \\ \hline \rule[0cm]{0cm}{.25cm}
$ m_{S_1} \geq 1.2\, M_{N} $  \\ \hline \rule[0cm]{0cm}{.25cm}
$  m_{S_1} \leq m_{S_2}\leq m_{S_3} \leq 2\,{\rm TeV} $   \\ \hline \rule[0cm]{0cm}{.25cm}
$ 10^2\,\mathrm{GeV}<\lambda_1v_{\sigma},\lambda_2 v_{\sigma} <10^3\, \mathrm{GeV} $  \\ \hline \rule[0cm]{0cm}{.35cm}
$10^{-6}\leq f_{i \alpha}, \kappa_1,\kappa_2 \leq 1$ \\ \hline \rule[0cm]{0cm}{.25cm}
$10^9\,{\rm GeV}\leq v_{\sigma} \leq 10^{13}\,{\rm GeV}$ \\ \hline 
\hline
 \end{tabular}
 \end{center}
 \caption{Random sampling for the relevant free parameters used in the numerical analysis.
 }
 \label{tab:parameterscan}
 \end{table}

\section{Results and Discussion}
\label{sec:results}

To study the fermion DM phenomenology in this model, we scanned the free parameters of the model as shown in Table \ref{tab:parameterscan}. We assumed $\lambda_{\sigma} = \lambda_{S_\alpha} = \lambda_{HS_{\alpha}} = 10^{-4}$, with $\left(\alpha=1,2,3\right)$\footnote{We implemented the model in  {\tt SARAH} \cite{Staub:2013tta, Staub:2015kfa} that calculates, via {\tt SPheno} \cite{Porod:2003um, Porod:2011nf} and {\tt FlavorKit} \cite{Porod:2014xia},  the mass spectrum, the oblique parameters and the flavor observables. Also, we used {\tt micrOMEGAs} \cite{Belanger:2013oya} to calculate the WIMP relic abundance.}. Moreover,  the mass of the exotic quark, $M_Q$ is set to $M_Q\sim 10$ TeV along with $y_Q = 0.1$ to stay safe from LHC constraints \cite{Alves:2016bib}. Let us recall that the Yukawa couplings $h_{i \alpha}$ are related to the Yukawa couplings $f_{i \alpha}$, neutrino masses, and the PNMS mixing matrix elements, as shown in section \ref{sec:neutrinomasses}. We guarantee that the charged LFV observables remain within the current experimental bounds\footnote{ For a summary on the current limits and the prospects experimental settings on charged LFV observables, see Table III in Ref. \cite{Carvajal:2021fxu}.}. Regarding neutrino physics, we consider NH for neutrino masses and use the best-fit point values reported in Refs. \cite{deSalas:2017kay, deSalas:2020pgw} for the $\mathcal{CP}$ conserving case.
\begin{figure}[t!]
  \includegraphics[scale=0.6]{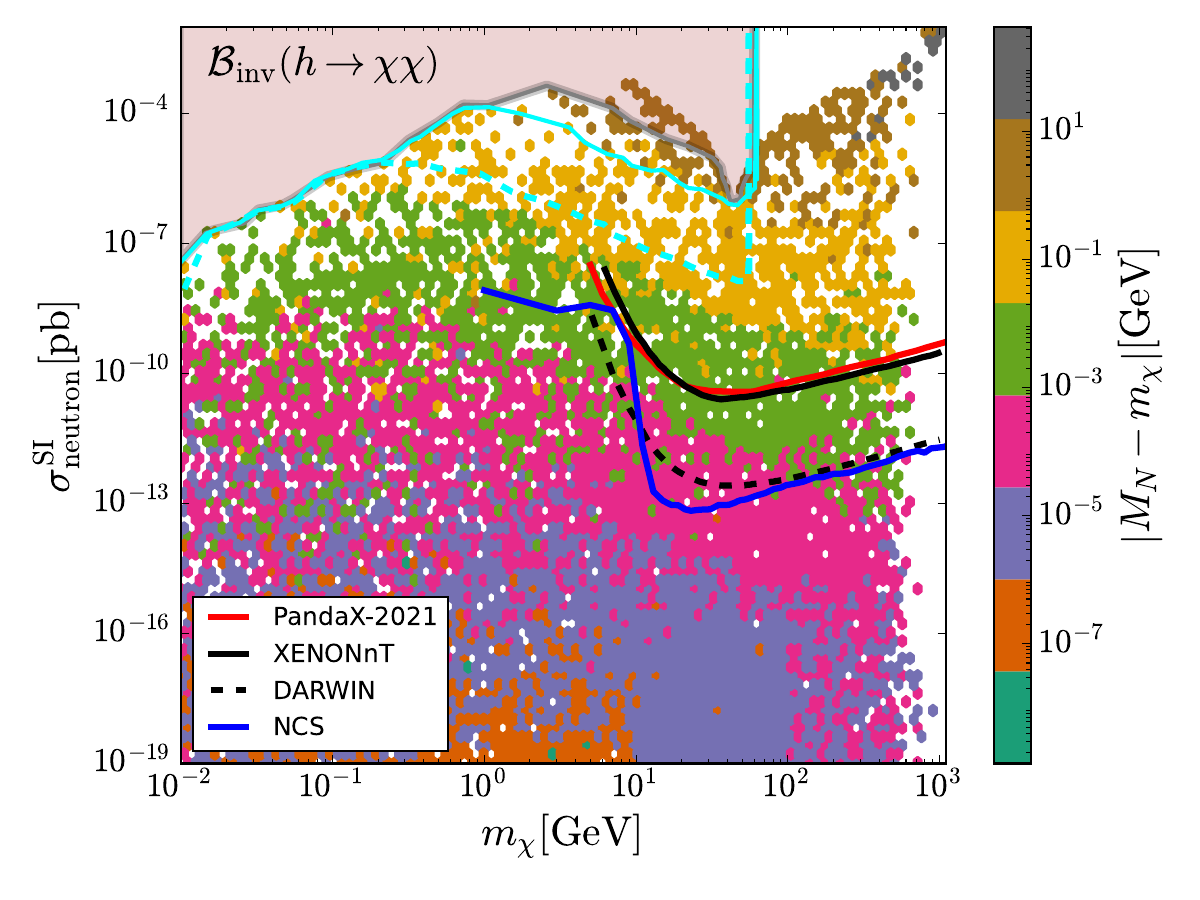}
  \caption{DM spin-independent cross sections as a function of the DM mass. The shaded region is excluded by the invisible Higgs decay into DM pair, $\mathcal{B}\left(h\rightarrow \chi \chi\right)$ \cite{ATLAS:2020cjb}, while the continuous and dashed cyan lines are the prospect limits expected from high luminosity LHC and Future Circular Colliders (FCC) \cite{deBlas:2019rxi}. }
  \label{fig:cross_section}
\end{figure}

\begin{figure}[t!]
  \includegraphics[scale=0.6]{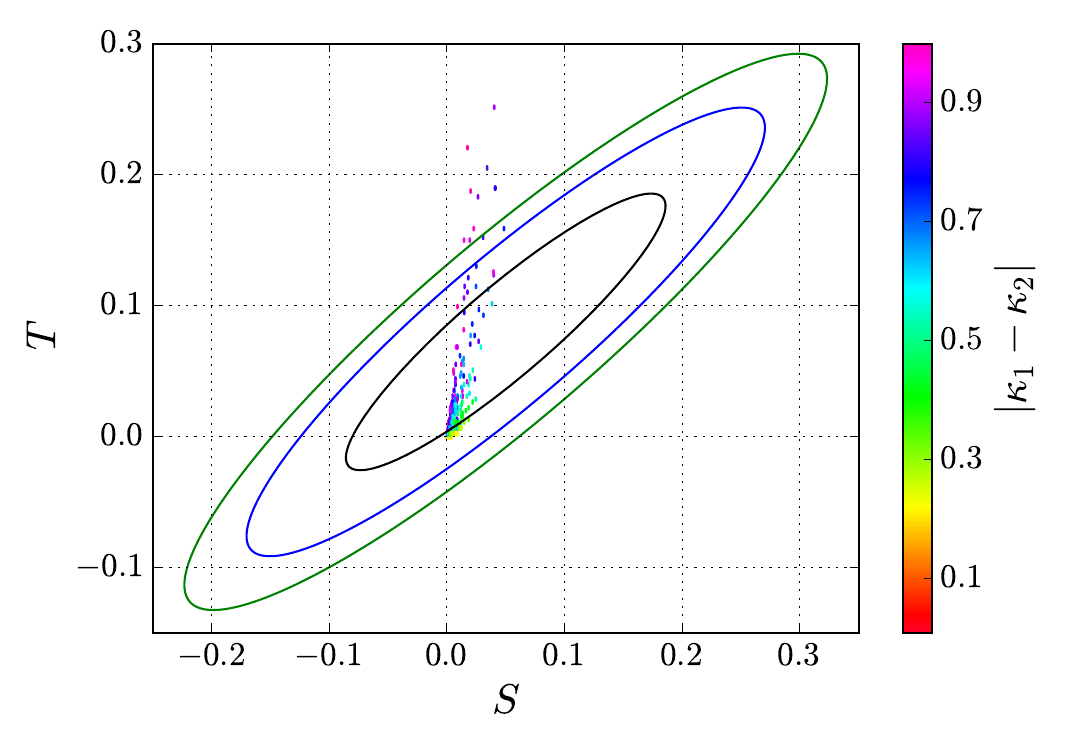}
  \caption{ Contour plot for scalar and fermion contributions to the EWPT parameters. The color code represents the mixing between the singlet and the doublet neutral fermion states. The
black, blue, and green ellipses represent the experimental constraints at 68\% CL, 95\% CL, and 99\% CL, respectively \cite{Baak:2014ora}.}
  \label{fig:STU}
\end{figure}

\begin{figure}[t!]
  \includegraphics[scale=0.6]{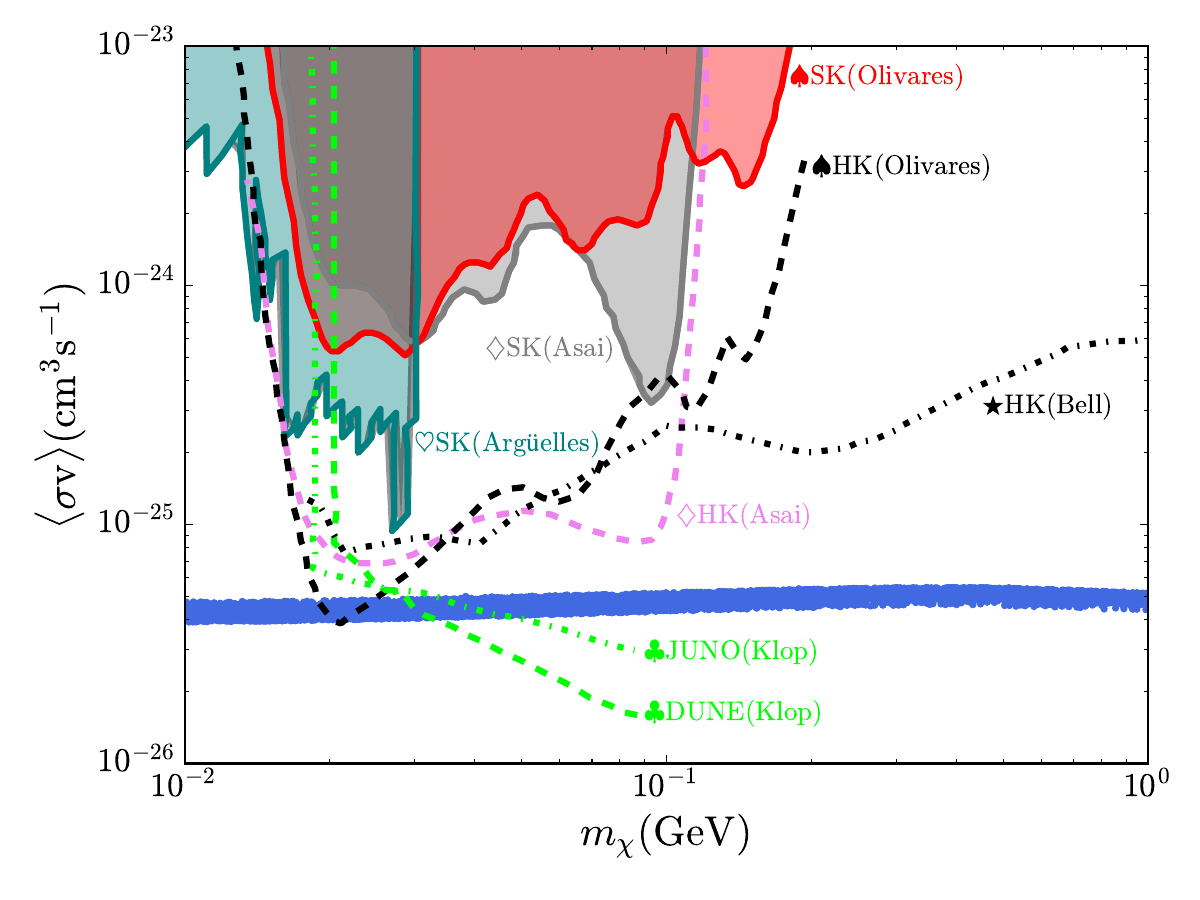}
  \caption{Thermally averaged DM annihilation cross section times velocity as a function of the DM mass and search limits imposed by DM annihilation into neutrinos in the Milky Way galaxy~\cite{Arguelles:2019ouk}.}
  \label{fig:indirectD}
\end{figure}

Results are shown in Fig. \ref{fig:cross_section}, Fig. \ref{fig:STU}, and Fig. \ref{fig:indirectD} which assume a total amount of WIMP DM component. Each point reproduces the observed DM relic density $\Omega h^2 = 0.120\pm0.001$ at $3\sigma$ \cite{Planck:2018vyg} and satisfies the current charged LFV bounds. Fig. \ref{fig:cross_section}  shows the direct detection cross section as a function of the DM mass. The color code represents the difference between the singlet and doublet fermion masses. The shaded region is excluded by the constraint of invisible Higgs decay into DM pair, $\mathcal{B}\left(h\rightarrow \chi \chi\right)$ \cite{ATLAS:2020cjb}, while the continuous and dashed cyan lines are the prospect limits expected from high luminosity LHC and Future Circular Colliders (FCC) \cite{deBlas:2019rxi}. Moreover, we plot the direct detection current limits imposed by XENONnT (solid black line) \cite{XENON:2023sxq} and PandaX-2021 (solid red line) \cite{PandaX-4T:2021bab} as well as the prospects that are expected from DARWIN (dashed black line) \cite{DARWIN:2016hyl}. The blue line represents the coherent elastic neutrino scattering (neutrino floor)~\cite{Graf:2013cqj, Billard:2013qya}. In Fig. \ref{fig:cross_section}, a sizeable mixing between the singlet and the doublet states gives a small vector coupling to the $Z$ boson. It is currently excluded from direct detection searches. Consequently, a small mixing, $\left|\kappa_1-\kappa_2\right|\lesssim 10^{-4}$, is required for DM masses $m_{\chi}\gtrsim$ 10 GeV. Such small Yukawa couplings also guarantee that the contributions from the new fermions to the oblique parameters remain at $3\sigma$ level \cite{Baak:2014ora}, as we show in Fig. \ref{fig:STU}. This limit on the Yukawa parameters $(\kappa_1,\kappa_2)$ is equivalent to an upper limit on the difference between the $M_N$ parameter and the DM mass, $\left|M_N - m_{\chi}\right|\lesssim 10^{-3}$ GeV, {\it i.e}, the DM state should be mostly singlet, $\chi = N$. Also, from Fig~\ref{fig:cross_section}, notice that the region of parameter space for the mass of the DM candidate below $100$ GeV is recovered.  

Fig. \ref{fig:indirectD} shows the DM annihilation cross section into a neutrino pair as a function of the DM mass, $m_{\chi}$.  We follow the notation in \cite{Arguelles:2019ouk} and  \cite{Okawa:2020jea}. The shadow region is the current exclusions extracted from Olivares ($\spadesuit$), \cite{Olivares-DelCampo:2017feq}, Asai ($\lozenge$) \cite{Asai:2020qlp} and Argüelles ($\heartsuit$) \cite{Arguelles:2019ouk}. Whereas, the dash, dotted, and dot-dashed represent the future searches reported by Olivares \cite{Olivares-DelCampo:2018pdl}, Bell ($\bigstar$) \cite{Bell:2020rkw}  and Klop ($\clubsuit$) \cite{Klop:2018ltd}.

In Fig. \ref{fig:indirectD} the model reaches $ \langle \sigma v \rangle \sim 4\times 10^{-26}$ cm$^3$ s$^{-1}$ which corresponds to the canonical thermally averaged cross-section. Notice that the future sensitivity of neutrino telescopes could test this model for DM masses  20 MeV $ \lesssim m_{\chi} \lesssim $ 30 MeV using HK searches and 30 MeV $ \lesssim m_{\chi} \lesssim $ 50 MeV using JUNO and DUNE combinations. Moreover, we emphasize that constraints from current DD searches do not affect that region of the parameter space. 

A final comment regarding the exotic quark $D$. Since it couples to the SM sector through the Yukawa term, $\overline{D_L}d_RS_{\alpha}$, it can decay into a scalar $S_{\alpha}$ and an SM quark, thus avoiding potential issues that arise when an exotic quark is considered cosmologically stable \cite{Nardi:1990ku}.

\section{Conclusions}
\label{sec:conclusions}

We analyze phenomenologically a model where the PQ mechanism is the solution to the strong CP problem and generates a Dirac mass term for neutrinos. Besides, a remnant $Z_2$ symmetry guarantees the existence of a WIMP DM candidate in addition to the axion. The WIMP DM phenomenology resembles the well-known singlet-doublet fermion DM, but in our case, we show that the low mass regime is recovered because of the new annihilation channels that result from the PQ mechanism that allow DM annihilation into neutrinos. This model solves the DM problem and generates neutrino masses and it could be tested in future experiments of DD of DM. Furthermore, we demonstrate that future neutrino telescopes will test the MeV region for DM masses where DD searches are not sensitive.

\section*{Acknowledgments}

We want to thank Walter Tangarife for very valuable feedback in the course of this work. 
The work of David Suarez and Robinson Longas is supported by Sostenibilidad UdeA,
UdeA/CODI Grant 2020-33177, and Minciencias Grants CD 82315 CT ICETEX 2021-1080 and 80740-492-2021. 

\appendix

\section{Texture of the Yukawa couplings involved in neutrino physics. }
\label{sec:Yukawas}

In several Majorana neutrino models, the Yukawa parameters are related to the neutrino physics via the Casas-Ibarra parametrization \cite{Casas:2001sr}. A generalization of the Casas-Ibarra parametrization is called the master equation and was studied in Ref. \cite{Cordero-Carrion:2019qtu} and can be used in many Majorana neutrino mass models. Following the motivation presented in Ref. \cite{Cordero-Carrion:2019qtu}, we study in this section a general solution for the Dirac neutrino mass couplings presented in Eq. \eqref{eq:neutrinomasses}. 

The mass matrix from any Dirac neutrino model can be written in the form: 
\begin{align}\label{eq:My1y2}
 M^{\nu}   = y_1 \Lambda y_2 \,,
\end{align}
where $\Lambda$ is a $3\times 3$ complex matrix with dimension of mass and the Yukawa couplings $y_1, y_2$, are dimensionless $3 \times 3$ complex matrices. We assume $\Lambda $ in the diagonal basis\footnote{A general analysis for an arbitrary dimension of $y_1, y_2$ and $\Lambda$ matrices will be left for future work.}. Note that the mass matrix structure in Eq. \eqref{eq:My1y2} contains several Dirac neutrino models and the results derived here can be applied to those models.
On the other hand, the data coming from neutrino oscillation requires at least two non-zero eigenvalues for the $M^{\nu}$ matrix. Then, the neutrino mass matrix must be ${\rm rank}\left(M^{\nu}\right) \equiv {\rm rank}\,{\rm M}   =  $ 2 or ${\rm rank}\,{\rm M} =$ 3. 
We will study both cases: NH and IH. 

\subsection{rank(M) = 2}

In the case of two non-zero neutrino mass eigenstates, the mass matrix in Eq.\eqref{eq:My1y2} can be diagonalized by a biunitary transformation, 
\begin{align}\label{eq:diagonal1}
m = U^{\dagger} M^{\nu} V = U^{\dagger} y_1\Lambda y_2 V = \left\{ \begin{array}{lcc}
{\rm diag}(0,m_2,m_3) &   {\mathrm{for}}  & {\mathrm{NH}}\,,  
\\ {\rm diag}(m_1,m_2,0) &   {\mathrm{for}}   & {\mathrm{IH}}\,. 
\end{array}
\right.
\end{align}
After that, we follow the same strategy presented in Ref. \cite{Cordero-Carrion:2019qtu} and we define the matrices: 
\begin{align}\label{eq:definitionD}
\sqrt{D} = \left\{ \begin{array}{lcc}
{\rm diag}(\sqrt{v},\sqrt{m_2},\sqrt{m_3}) &   {\mathrm{for}}  & {\mathrm{NH}}  \,,
\\ {\rm diag}(\sqrt{m_1},\sqrt{m_2},\sqrt{v}) &   {\mathrm{for}}   & {\mathrm{IH}} \,,
\end{array}
\right.
\end{align}
where $v$ is some non-vanishing arbitrary energy scale. For example, in this work $v\equiv v_{\sigma}$. It is worth mentioning that the analytical expressions found for the Yukawa couplings are independent of the choice of the energy scale $v$ as it was discussed in \cite{Cordero-Carrion:2019qtu}.

If we multiply the expression in Eq. \eqref{eq:diagonal1} on the left and on the right side by $\sqrt{D}^{-1}$, we obtain: 
\begin{align}\label{eq:diagonal2}
\sqrt{D}^{-1} \sqrt{m}\sqrt{m}\sqrt{D}^{-1} \equiv R = \sqrt{D}^{-1}U^{\dagger} y_1\Lambda y_2 V\sqrt{D}^{-1} \,,
\end{align}
where we use the definitions for $m$ and $\sqrt{D}$ to write the left side of the equation. With this, the matrix $R$ is defined by:
\begin{align}\label{eq:definitionR}
    R = \left\{ \begin{array}{lcc}
\begin{pmatrix}
0&0&0\\
0&1&0\\
0&0&1
\end{pmatrix} &   {\mathrm{for}}  & {\mathrm{NH}}  \;,
\\ \\ \begin{pmatrix}
1&0&0\\
0&1&0\\
0&0&0
\end{pmatrix} &   {\mathrm{for}}   & {\mathrm{IH}} \;.
\end{array}
\right.
\end{align}

The expression in Eq. \eqref{eq:diagonal2} can be written in the form:
\begin{align}\label{eq:diagonal4}
 &R = \sqrt{D}^{-1}U^{\dagger} y_1\sqrt{\Lambda}\sqrt{\Lambda} y_2 V\sqrt{D}^{-1}  \\
 &R = \left[ \sqrt{\Lambda}y_1^{\dagger}U\sqrt{D}^{-1} \right]^{\dagger}\left[\sqrt{\Lambda} y_2 V\sqrt{D}^{-1}\right] \label{eq:diagonal5} \\
&R \equiv R_1^{\dagger}R_2\;,\label{eq:diagonal6}
\end{align}
where we use the fact that $\Lambda$ is a diagonal matrix and we define the matrices $R_1$ and $R_2$ as,
\begin{align}\label{eq:definitionR1R2}
    &R_1 =  \sqrt{\Lambda}y_1^{\dagger}U\sqrt{D}^{-1} \,,\\
    &R_2 = \sqrt{\Lambda} y_2 V\sqrt{D}^{-1}\;.
    \label{eq:definitionR1R22}
\end{align}

The existence of an inverse for the matrices $R_1$ and $R_2$ in Eq.\eqref{eq:diagonal6} allows us to express the Yukawa coupling $y_1$ ($y_2$) as a function of the Yukawa coupling $y_2$ ($y_1$) and the neutrino oscillation observables. Alternatively, either $y_1$ or $y_2$ remains as a free parameter in the model as follow\footnote{In Ref. \cite{Guo:2020qin} similar formulas are reported. The authors use a general parametrization for the matrices $R_1, R_2$ and the relation in Eq. \eqref{eq:diagonal6}. Here, conversely, we choose to write one Yukawa coupling in terms of the other one.}: if we multiply Eq. \eqref{eq:diagonal6} at the right by $R_2^{-1}$, we extract $y_1$ by using Eqs.\eqref{eq:definitionR1R2}, \eqref{eq:definitionR1R22} and we obtain:
\begin{align}\label{eq:yukawageneralr2y1}
    y_1 = U \sqrt{D}R\sqrt{D}V^{\dagger}y_2^{-1}\Lambda^{-1}\;.
\end{align}
Conversely, if we multiply Eq.\eqref{eq:definitionR1R2} at the left by $\left(R_1^{\dagger}\right)^{-1}$, we extract $y_1$ by using Eqs.\eqref{eq:definitionR1R2}, \eqref{eq:definitionR1R22} and obtain,
\begin{align}\label{eq:yukawageneralr2}
    y_2=\Lambda^{-1}y_1^{-1}U\sqrt{D}R\sqrt{D}V^{\dagger}\;.
\end{align}
For the case of the model we study here, the result in Eq. \eqref{eq:yukawaf} is obtained as particular case of  Eq. \eqref{eq:yukawageneralr2y1} by setting $V=\mathbb{1}$, $U=U_{{\rm PMNS}}$, $y_1 = h$ and $y_2 = f^T$. In short:
\begin{align}\label{eq:yukawageneralr3}
     h=U_{{\rm PMNS}}\sqrt{D}R\sqrt{D}\left(f^T\right)^{-1}\Lambda^{-1}\;.
\end{align}

\subsection{rank(M) = 3}

In the case of three neutrinos being massive, we follow the same method. The mass matrix in Eq.\eqref{eq:My1y2} can be diagonalized by a biunitary transformation, 
\begin{align}
    &m = {\rm diag}(m_1,m_2,m_3) = U^{\dagger} y_1 \sqrt{\Lambda}\sqrt{\Lambda}y_2 V^{\dagger}\;, \\
    & \sqrt{m}\sqrt{m} = U^{\dagger} y_1 \sqrt{\Lambda}\sqrt{\Lambda}y_2 V^{\dagger}\;. \label{eq:diagonal7}
\end{align}
We then multiply the last expression at the right side and the left side by $\sqrt{m}^{-1}$ and obtain,
\begin{align}
\label{eq:yukarank3-1}
    \sqrt{m}^{-1}\sqrt{m}\sqrt{m}\sqrt{m}^{-1} = \mathbb{1}_3
    = \sqrt{m}^{-1} U^{\dagger} y_1 \sqrt{\Lambda}\sqrt{\Lambda}y_2 V^{\dagger}\sqrt{m}^{-1}  \;,
\end{align}
where $\mathbb{1}_3$ is the $3\times 3 $ identity matrix. The expression in Eq. \eqref{eq:yukarank3-1} is reorganized in the form:
\begin{align}
    &\left[\sqrt{\Lambda} y_1^{\dagger} U \sqrt{m}^{-1}\right]^{\dagger}\left[\sqrt{\Lambda}y_2 V^{\dagger}\sqrt{m}^{-1} \right] = \mathbb{1}_3\;, \\
    & R_1^{\dagger}R_2 =\mathbb{1}_3\;.\label{eq:diagonal8}
\end{align}
Again, the existence of an inverse for the matrices $R_1$ and $R_2$ in Eq. \eqref{eq:diagonal8} allows us to set one of the Yukawa couplings in terms of the other one and the neutrino observable parameters as,
\begin{align}
    y_1 = U m V^{\dagger}y_2^{-1}\Lambda^{-1}\;\;{\text{or}}\;\; y_2=\Lambda^{-1}y_1^{-1}UmV^{\dagger} \;.
\end{align}
The identifications $V=\mathbb{1}$, $U=U_{{\rm PMNS}}$, $y_1 = h$ and $y_2 = f^T$ in the previous equation describe the Yukawa texture of the model presented here. 

\bibliographystyle{apsrev4-1long}
\bibliography{references}

\end{document}